\documentclass{icrc2009}

\usepackage{graphicx}   
\usepackage[caption=false]{caption}    
\usepackage[font=footnotesize]{subfig} 
\usepackage{fixltx2e}
\usepackage{url}

\usepackage{cite}

\newcommand{\shorttitle}[1]%
{\markboth{Proceedings of the 31\MakeLowercase{$^{st}$} ICRC, {\L}\'{o}d\'{z} 2009}{#1} }
\newcommand{\etal}{\MakeLowercase{\textit{et al. }}} 

\newcommand{\ej}{\epsilon_{3e51}^{\rm jet}}
\newcommand{\rhosn}{\rho_{2e-4}^{\rm SN}}
\newcommand{\nab}{N^{\rm SN,AB}_{\rm IC/ROTSE}}
\newcommand{\nalerts}{N_{\rm alerts}}
\newcommand{\deltat}{\Delta t_d}
\newcommand{\ntotal}{N_{\rm ROTSE}}

\newcommand{\gsim}{\mathrel{\rlap{\lower4pt\hbox{\hskip1pt$\sim$}}\raise1pt\hbox{$>$}}}               

\begin{document}
\title{Optical follow-up of high-energy neutrinos detected by IceCube}

\author{\IEEEauthorblockN{Anna Franckowiak\IEEEauthorrefmark{1},
                          Carl Akerlof\IEEEauthorrefmark{4},
			  D.~F.~Cowen\IEEEauthorrefmark{1}\IEEEauthorrefmark{2},
                          Marek Kowalski\IEEEauthorrefmark{1},
                          Ringo Lehmann\IEEEauthorrefmark{1},\\
                          Torsten Schmidt\IEEEauthorrefmark{3} and
                          Fang Yuan\IEEEauthorrefmark{4}} for the IceCube Collaboration\IEEEauthorrefmark{5} and for the ROTSE Collaboration
                            \\
\IEEEauthorblockA{\IEEEauthorrefmark{1}Humboldt Universit\"at zu Berlin}
\IEEEauthorblockA{\IEEEauthorrefmark{2}Pennsylvania State University}
\IEEEauthorblockA{\IEEEauthorrefmark{3}University of Maryland}
\IEEEauthorblockA{\IEEEauthorrefmark{4}University of Michigan}
\IEEEauthorblockA{\IEEEauthorrefmark{5}see http://www.icecube.wisc.edu/collaboration/authorlists/2009/4.html}}

\shorttitle{Author \etal paper short title}
\maketitle

\begin{abstract}
Three-quarters of the 1~km$^3$ neutrino telescope IceCube is currently taking data. Current models predict high-energy neutrino emission from transient objects like supernovae (SNe) and gamma-ray bursts (GRBs). To increase the sensitivity to such transient objects we have set up an optical follow-up program that triggers optical observations on multiplets of high-energy muon-neutrinos. We define multiplets as a minimum of two muon-neutrinos from the same direction (within 4$^{\circ}$) that arrive within a 100~s time window. When this happens, an alert is issued to the four ROTSE-III telescopes, which immediately observe the corresponding region in the sky. Image subtraction is applied to the optical data to find transient objects. In addition, neutrino multiplets are investigated online for temporal and directional coincidence with gamma-ray satellite observations issued over the Gamma-Ray Burst Coordinate Network. An overview of the full program is given, from the online selection of neutrino events to the automated follow-up, and the resulting sensitivity to transient neutrino sources is presented for the first time.
  \end{abstract}

 
\section{Introduction}
\begin{figure}[t!]
\centering
\includegraphics[width=3.8in]{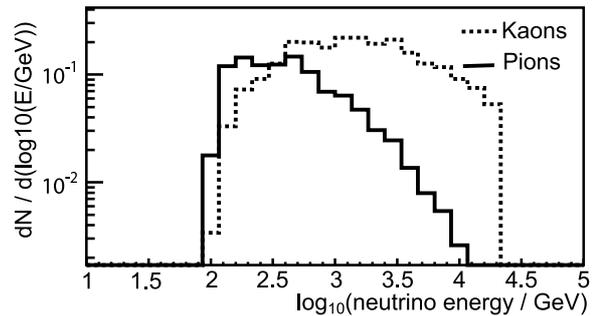}
\caption{Neutrino event spectrum in the IceCube detector, from kaon and pion decay in the supernovae-jet model of Ando and Beacom \cite{AndoBeacom}.}
\label{fig:spectrum}
\end{figure}

When completed, the in-ice component of IceCube will consist of 4800 digital optical modules (DOMs) arranged on 80 strings frozen into the ice, at depths ranging from 1450m to 2450m~\cite{IceCube}. Furthermore there will be six additional strings densely spaced at the bottom half of the detector. The total instrumented volume of IceCube will be 1~km$^3$. Each DOM contains a photomultiplier tube and supporting hardware inside a glass pressure sphere. The DOMs indirectly detect neutrinos by measuring the Cherenkov light from secondary charged particles produced in neutrino-nucleon interactions. IceCube is most sensitive to neutrinos within an energy range of TeV to PeV and is able to reconstruct the  direction of muon-neutrinos with a precision of $\sim$1$^{\circ}$.\\
The search for neutrinos of astrophysical origin is among the primary goals of the IceCube neutrino telescope. Source candidates include galactic objects like supernova remnants as well as extragalactic objects like Active Galactic Nuclei and Gamma-Ray Bursts~\cite{Julia}~\cite{HalzenHooper}. Offline searches for neutrinos in coincidence with GRBs have been performed on AMANDA and IceCube data. They did not lead to a detection yet, but set upper limits to the predicted neutrino flux~\cite{Kyler}. While the rate of GRBs with ultra-relativistic jets is small, a much larger fraction of SNe not associated with GRBs could contain mildly relativistic jets. Such mildly relativistic jets would become stalled in the outer layers of the progenitor star, leading to essentially full absorption of the electromagnetic radiation emitted by the jet. Hence, with the postulated presence of mildly relativistic jets one is confronted with a plausible but difficult-to-test hypothesis. Neutrinos may reveal the connection between GRBs, SNe and relativistic jets. As was recently shown, mildly relativistic jets plowing through a star would be highly efficient in producing high-energy neutrinos \cite{AndoBeacom, Razzaque, Horiuchi}. The predicted neutrino spectrum follows a broken power law and Fig.~\ref{fig:spectrum} shows the expected signal spectrum for neutrinos produced in kaon and pion decay in the source, simulated using the full IceCube simulation chain.
The expected number of signal events is small and requires efficient search algorithms to reduce the background of atmospheric neutrinos (see section~\ref{sec:EventSelection}).\\
An optical follow-up program has been started which enhances the sensitivity for detecting high-energy neutrinos from transient sources such as SNe. In this program, the direction of neutrinos are reconstructed online, and if their multiplicity pass a certain threshold, a Target-of-Opportunity (ToO) notice is sent to the ROTSE-III network of robotic telescopes. These telescopes monitor the corresponding part of the sky in the subsequent hours and days and identify possible transient objects, e.g. through detection of rising supernova light-curves lasting several days. If in this process a supernova is detected optically, one can extrapolate the lightcurve or afterglow to obtain the explosion time~\cite{T0paper}. For SNe, a  gain in sensitivity of about a factor of 2-3 can be achieved through optical follow-up observations of neutrino multiplets~\cite{MarekAnna}. In addition to the gain in sensitivity, the follow-up program offers a chance to identify its transient source, be it a SN, GRB or any other transient phenomenon.

\section{Neutrino Alert System}
\label{sec:EventSelection}
IceCube's optical follow-up program has been operating since fall of 2008. In order to match the requirements given by limited observing time at the optical telescopes, the neutrino candidate selection has been optimized to obtain less than about 25 background multiplets per year. The trigger rate of the 40 string IceCube detector is about 1000~Hz. The muon filter stream reduces the rate of down-going muons created in cosmic ray showers dramatically by limiting the search region to the Northern hemisphere and a narrow belt around the horizon. The resulting event stream of 25~Hz is still dominated by misreconstructed down-going muons. Selection criteria based on on track quality parameters, such as number of direct hits\footnote{Hits that are measured within $[$-15ns,75ns$]$ from the predicted arrival time of Cherenkov photons, without scattering, given by the track geometry.}, track length and likelihood of the reconstruction, yield a reduced event rate of 1~event/(10~min). The optimized selection criteria are relaxed to improve the signal efficiency, 50\% of the surviving events are still misreconstructed down-going muons, while 50\% are atmospheric neutrinos. During the antarctic summer 2008/2009, 19 additional strings were deployed, which have been included in the data taking since end of April 2009. To take into account an enhancement in the rate due to the increased detector volume, the selection criteria have been adjusted and will yield a cleaner event sample containing only 30\% misreconstructed muons.\\ 
From this improved event sample, neutrino multiplet candidates with a time difference of less than 100~s and with an angular difference (or 'space angle') of less than 4$^{\circ}$ are selected. The choice of the time window size is motivated by jet penetration times. Gamma-ray emission observed from GRBs has a typical length of 40~s, which roughly corresponds to the duration of a highly relativistic jet to penetrate the stellar envelope. The angular difference is determined by IceCube's angular resolution. Assuming single events from the same true direction, 75\% of all doublets are confined to a space angle of 4$^{\circ}$ after reconstruction.
Once a multiplet is found, a combined direction is calculated as a weighted average of the individual reconstructed event directions, with weights derived from the estimated direction resolution of each track.  The resolution of the combined direction is up to a factor of $1/\sqrt{2}$ better than that of individual tracks. 
The multiplet direction is sent via the network of Iridium satellites from the South Pole to the North, where it gets forwarded to the optical telescopes. At this point in time, due to limited parallelization of the data processing at the South Pole a delay of 8~hours is accumulated. In the near future, the online processing pipeline will be upgraded, 
reducing the latency drastically to the order of minutes.\\
A total of 14 alerts have passed the selection criteria and were sent to the telescopes within 7 months of operation.

\section{Optical Follow-Up Observations}
At the moment IceCube alerts get forwarded to the Robotic Optical Transient Search Experiment (ROTSE)~\cite{rotse}. Additions to the list of participating telescopes are planned. ROTSE-III is dedicated to observation and detection of optical transients on time scales of seconds to days. The original emphasis was on GRBs while it more recently has also started a very successful SN program. The four ROTSE-III telescopes are installed around the world (in Australia, Namibia, the USA and Turkey). The ROTSE-III equipment is modest by the standards of modern optical astronomy, but the wide field of view and the fast response permit measurements inaccessible to more conventional instruments. The four 0.45 m robotic reflecting telescopes are managed by a fully-automated system. They have a wide field of view (FOV) of $1.85^{\circ} \times 1.85^{\circ}$ imaged onto a 2048 $\times$ 2048 CCD, and operate without filters. The cameras have a fast readout cycle of 6~s. The limiting magnitude for a typical 60 s~exposure is around 18.5~mag, which is well-suited for a study of GRB afterglows during the first hour or longer. The typical full width at half maximum (FWHM) of the stellar images is smaller than 2.5 pixels (8.1 arcseconds). Note that ROTSE-III’s FOV matches the size of the point spread function of IceCube well.\\
Once an IceCube alert is received by one of the telescopes, the corresponding region of the night sky will be observed within seconds. A predefined observation program is started: The prompt observation includes thirty exposures of 60~seconds length~\footnote{Once the delay caused by data processing at the South Pole (see section~\ref{sec:EventSelection}) is reduced to the order of minutes, the prompt observation will include ten short observations of 5~seconds, ten observations of 20~seconds and twenty long exposures of 60~seconds.}. Follow-up observations are performed for 14 nights. Eight images with 60~seconds exposure time are taken per night. The prompt observation is adjusted to the typical rapidly decaying lightcurve of a GRB afterglow, while the follow-up observation of 14 days permits the identification of an increasing SN lightcurve. Once the images are taken, they are automatically processed at the telescope site. 
Once the data is copied from the telescopes, a second analysis is performed 
off-line, combining the images from all sites. Image subtraction is performed according the methods presented in~\cite{Subtraction}. Here the images of the first night serve as reference, while the images from the following nights are used to search for the brightening of a SN lightcurve.

\section{Sensitivity}
The sensitivity of the optical-follow up program is determined by both IceCube's sensitivity 
to high energy neutrino multiplets and ROTSE-III's sensitivity to SNe. We will distinguish two cases: The first being that no optical counterpart is observed
over the course of the program (assuming 25 alerts per year) and the second that a SN is identified in coincidence.
\subsection{No Optical Counterpart Discovered}
\begin{figure}[t!]
\centering
\includegraphics[width=3.4in]{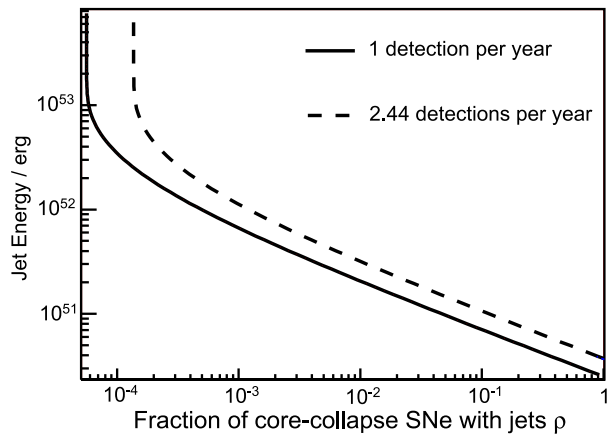}
\caption{Sensitivity in the $E_{\rm jet}-\rho$ plane after one year of
  operation of the 40 string IceCube detector (dashed line---90\% CL; solid line---one coincident
  detection per year).}
\label{fig:limit}
\end{figure}

With no coincident SN observed, one obtains an upper limit on the
average number of SNe that could produce a coincidence: $N_{\rm
  ROTSE}^{\rm IC}<2.44$ (for 90\% confidence level).  Constraints on a given model
are obtained by demanding that the model does not predict a number in
excess of the SN event upper limit. We construct a simple model based
on Ando \& Beacom type SNe~\cite{AndoBeacom}. We introduce two parameters: The first is
the rate of SNe producing neutrinos $\rho= (4/3 \pi)^{ -1} 10^{-4}
\rhosn {\rm Mpc}^{-3} {\rm yr}^{-1}$.  Note that $\rhosn=1$ corresponds to one
SN per year in a 10~Mpc sphere, about the rate of all core-collapse SNe in
the local Universe~\cite{SNRate}. Since we expect only a subset of SNe to produce
high energy emission, one can assume $\rhosn<1$.  The second parameter
is the hadronic jet energy $E_{\rm jet}= 3\cdot 10^{51} \ej \rm ergs$
and we choose to scale the flux normalization of the model of
Ando \& Beacom, $F_0$, by $\ej$.  Fig. \ref{fig:limit} shows the
constraints that one can place on the density and jet kinetic energy
in the $E_{\rm jet}-\rho$ plane.
The basic shape of the constraints that can be obtained in the $E_{\rm
jet}-\rho$ plane can be understood from the following
considerations. 
The number of neutrinos depends on the jet energy and the distance: $ N_{\nu} \propto \ej \cdot r^{-2}.$
The program requires at least $N_{\nu, \rm min}=2$ detected neutrinos in IceCube. A SN with jet energy $\ej$ produces $N_{\nu, \rm min}$ neutrinos if it is closer than $r_{\rm max}$:
$N_{\nu, \rm min} \propto \ej \cdot r_{\rm max}^{-2}\rm,$
which yields $r_{\rm max} \propto (\ej)^{1/2}$. The volume $V$ limited by $r_{\rm max}$ contains $N_{\rm SN} \propto \rhosn \cdot r_{\rm max}^{3}$ SN that can produce two neutrinos. Therefore the number of detection $N_{\rm IC/ROTSE}^{\rm SN}$ is given by $N_{\rm IC/ROTSE}^{\rm SN} \propto \rhosn \cdot (\ej)^{3/2}$.
For normalization we use Ando \& Beacom-like SNe, which occur at a rate of $\rhosn=1$ with GRB-like energies ($\ej=1$) and yield $\nab=200$ expected IceCube/ROTSE coincidences per year. 
\begin{equation}
N_{\rm IC/ROTSE}^{\rm SN} = \nab \rhosn \cdot (\ej)^{3/2}
\label{NSN}
\end{equation}
According to \cite{FeldmanCousins} a non-detection limits the number of IceCube/ROTSE coincidences at a 90\% confidence level to $N_{\rm IC/ROTSE}^{SN}<2.44$. Using Eq.~\ref{NSN} one obtains the two-dimensional constraints on density and hadronic jet energy for this model:
\begin{equation}
\rhosn (\ej)^{3/2} < 2.44/ \nab < 0.012,
\label{eq:upperlimit}
\end{equation}
which is a reasonably good representation of the two dimensional
constraints for not too small densities $\rhosn \gsim 10^{-3}$.  For
GRB-like energies ($\ej=1$), it follows that at most one out of 80
SNe produces Ando \& Beacom-like jets in its core. Phrased in
absolute terms, if no SN will be detected, the rate of SNe with a mildly relativistic jet should not
exceed $\rho = 3.1 \cdot 10^{-6} {\rm Mpc}^{-3}\rm{yr}^{-1}$ (at 90\% confidence level) in our program.
The cut-off at small densities visible in Fig. \ref{fig:limit} is due to ROTSE-III's limiting magnitude.
The sphere (i.e. effective volume) within which ROTSE-III can detect Supernovae has a radius of about 200-300 Mpc. ROTSE-III effectively cannot probe SN subclasses that occur less then once per year within this sphere.

\subsection{Significance in case of a detection}
Next we address the case that a SN was detected in the follow-up
observations.  The task mainly consists of computing the significance
of the coincidence. We compute this for one year of data and 25
alerts.  Each alert leads to the observation of a $\Delta \Omega =
1.85^\circ\times 1.85^\circ = 3.4$ square degree field, hence over the
course of the year ROTSE-III covers a fraction of the sky given by $\Delta \Omega /
4\pi \times \nalerts = 2.1 \cdot 10^{-3}$. Next assume that
the time window for a coincident of an optical SN detection and candidate neutrino multipet is given by $\deltat$, the accuracy with which we
can determine the initial time of the supernova explosion.
Studying the lightcurve of supernova SN2008D, which has a known explosion start-time given by an initial x-ray flash, we have developed an accurate way to estimate $\deltat$ from a SN lightcurve~\cite{T0paper}.
We fit the light curve data to a model that postulates a phase of
blackbody emission followed by a phase dominated by pure expansion of
the luminous shell. Explosion times can be determined from the lightcurve with an accuracy of less than 4 hours. A detailed description of this method can be found in~\cite{T0paper}.\\
The number of accidental SNe found will be
proportional to $\deltat$ and the total number of SNe per year that ROTSE-III would
have sensitivity to detect, if surveying the sky at all times, $\ntotal\approx10^4$. Putting all this together the number of
random coincidences is:
\begin{equation}
N_{\rm bg} = \nalerts \ntotal \frac{\Delta \Omega}{ 4\pi} \times
\frac{\deltat}{\rm yr} = 0.056 \frac{\deltat}{\rm d}.
\label{eq:rate_coincidence}
\end{equation}
For $N_{\rm bg}\ll 1$ this corresponds to the chance probability
$p=1-\exp(-N_{\rm bg})\approx N_{\rm bg}$ of observing at least one
random background event. For $\deltat=1 \rm d$ and no other
information, the observation of a SN in coincidence with a neutrino
signal would have a significance of about 2$\sigma$.\\
The significance can be improved by adding
neutrino timing information as well as the distance information of the
object found. We first discuss the extra timing information. So far
we have only required that two neutrinos arrive within 100\,s to
produce an alert.  Thus, in the analysis presented above, the
significance for two events 1\,s apart would be the same as for 99~s
difference. Since the probability $p_t$ to find a time difference less than
$\Delta t_\nu$ due to a background fluctuation is given by $p_t=\Delta
t_\nu / 100 s$ assuming a uniform background, we
include the time difference in the chance probability. 
Next we discuss the use of the SN distance. One can safely assume that there
will be a strong preference for nearer SNe, since these are most likely to
lead to a neutrino flux large enough to produce a multiplet in
IceCube. Using the distance $d_{\rm SN}$ as an additional parameter
one can compute the probability to observe a background SN at a
distance $d \le d_{\rm SN}$.  The probability is given by the ratio
of SNe observed by ROTSE-III within the sphere $d_{\rm SN}$ to all SNe:
$p_{d} = \ntotal(d)/\ntotal$.
In case of a detection both $d_{\rm SN}$ and $\Delta t_{\nu}$ will be available. We use a simple Monte Carlo to obtain the significance of this detection.
For example the detection of two neutrinos with a temporal difference of $\Delta t_\nu =$10~s in coincidence with a SN in $d_{\rm SN} =$20~Mpc distance has a p-value of $5\cdot 10^{-4}$, which corresponds to $3.5\sigma$, assuming a total of $\nalerts=25$ alerts found in the period of one year.

\section{Coincidences with GCN-GRBs} 

According to current models, about every 15-20th GRB that can be
detected by IceCube will produce a neutrino doublet. Hence there is a
small possibility that we will find a doublet in coincidence with a
GCN alert, a case that we consider separately here.  The significance
of such a coincidence can be estimated with calculation analogous
to Eq.~\ref{eq:rate_coincidence}. The number of accidental coincidences with a time difference less than $\Delta t$ is given by:
\begin{equation}
N_{\rm bg} = \nalerts N_{\rm GCN} \frac{\Delta \Omega}{ 4\pi} \times
\frac{\Delta t}{\rm yr} = 3.2 \cdot 10^{-8}\frac{\Delta t}{\rm 1 s,}.
\label{eq:NBG}
\end{equation}
where we have assumed 200 GCN notices and 30 multiplets a year.  A
coincidence occurs whenever the neutrinos and the GRB overlap within
predefined windows in direction and time.  For illustrative purposes,
if we choose a 1.5-degree directional window and a 4-hour time window
(corresponding roughly to IceCube's point spread function and to GRB
observations and modeling), Eq.~\ref{eq:NBG} yields an expected background count
of $N_{BG}=4.7\cdot 10^{-4}$. This corresponds to a $3.5\sigma$ effect, or
equivalently the expectation of a false positive from background once
every 2100 years.
We can further reduce the expected background by assuming that the
neutrino signal is most likely to be emitted at the same time as the
gamma rays.  Since the background multiplets will
be distributed uniformly across the 4-hour window, we can multiply the
chance probability above by the factor 
\begin{equation}
p_{t}=\left | \frac{t_{GRB}-t_{\nu}}{4\textrm{~hours}} \right |
\label{eq:pt_factor}
\end{equation}
where the absolute value is taken since we assume the neutrinos are
equally likely to be emitted before the gamma-rays as they are after.
Note that our flat probability assumption for the relative emission
times of gamma rays and neutrinos from GRBs can, of course, be
modified to follow any particular theoretical model.  With all these
assumptions, if we observe a coincidence that is 300 seconds
from the GRB onset time, the chance probability is then given by $N_{BG}\cdot p_{t} = 4.7\cdot 10^{-4} \cdot 300/14400 = 9.8\cdot 10^{-6}$, which corresponds to a $4.4\sigma$
result.
 
\section{Conclusion}
We have presented the setup and performance of IceCube's optical follow-up program, which was started in October 2008. The program increases IceCube's sensitivity to transient sources such as SNe and GRBs and furthermore allows the immediate identification of the source. Non-detection of an optical counterpart allows the calculation of a limit on model parameters such as jet energy and density of SN accompanied by jets.\\
In addition multiplets of neutrinos are tested for coincidences with GCN messages. Even a single coincidence detection would be significant.

\section{Acknowledgments}
A.~Franckowiak and M.~Kowalski acknowledge the support of the DFG. D.~F.~Cowen thanks the Deutscher Akademischer Austausch Dienst
(DAAD) Visiting Researcher Program and the Fulbright Scholar Program.


\begin{thebibliography}{99}
\bibitem{IceCube} A.~Achterberg et al. [IceCube Collaboration], Astropart.Phys. 26:155-173, 2006
\bibitem{T0paper}D.~F.~Cowen, A.~Franckowiak and M.~Kowalski, arXiv:0901.4877
\bibitem{rotse}C.~W.~Akerlof {\it et al.}, PASP 115:132-140, 2003
\bibitem{MarekAnna}M.~Kowalski, A.~Mohr, Astropart.Phys. 27:533-538,2007
\bibitem{AndoBeacom}S.~Ando, J.~Beacom, Phys.Rev.Lett. 95:061103,2005
\bibitem{Razzaque}S.~Razzaque, P.~Meszaros, E.~Waxman, Phys.Rev.Lett. 93:181101, 2004
\bibitem{Horiuchi}S.~Horiuchi, S.~Ando, Phys.Rev. D77:063007,2008. 
\bibitem{Subtraction}F.~Yuan, C.~W.~Akerlof, Astropart.Phys. 677:808-812,2008
\bibitem{Julia}J.~Becker, Phys.Rept. 458:173-246,2008 
\bibitem{HalzenHooper}F.~Halzen, D.~Hooper, Rept.Prog.Phys. 65:1025-1078, 2002 
\bibitem{FeldmanCousins} G.~J.~Feldman, R.~D.~Cousins, Phys.Rev. D57:3873-3889,1998
\bibitem{SNRate} S.~Ando, F.~Beacom, H.~Yuksel, Phys.Rev.Lett. 95:171101, 2005
\bibitem{Kyler} A.~Achterberg et al. [IceCube Collaboration], APJ 674:357-370, 2007



 \end{thebibliography}
\end{document}